\documentclass[preprint,eqsecnum,aps,prd,nofootinbib]{revtex4}
\usepackage[dvips]{graphicx}
\usepackage{graphics, color}
\usepackage{srcltx}
\usepackage{amsmath}
\usepackage{enumerate}
\usepackage{mathrsfs}
\usepackage{amsfonts}

 \newcommand{\be}{\begin{equation}}
\newcommand{\ee}{\end{equation}}

\newcommand{\Op}{{\mathcal O}}

\newcommand{\bbR}{{\mathbb{R}}}

\newcommand{\M}{{\mathcal M}}

\newcommand{\tr}{{\mathrm{tr}}}

\usepackage{hyperref}

\begin{document}

\title{Time evolution of entanglement entropy from a pulse}
\author{Matthew M.~Roberts}
\email{matthew.roberts@nyu.edu}
\affiliation{Center for Cosmology and Particle Physics, Department of Physics, New York University, 4 Washington Place, New York, NY 10003}

\begin{abstract} 
We calculate the time evolution of the entanglement entropy in a 1+1 CFT with a holographic dual when there is a localized left-moving packet of energy density. We find the gravity result agrees with a field theory result derived from the transformation properties of R\'enyi entropy. We are able to reproduce behavior which qualitatively agrees with CFT results of entanglement entropy of a system subjected to a local quench. In doing so we construct a finite diffeomorphism which tales three-dimensional anti-de Sitter space in the Poincar\'e patch  to a general solution, generalizing the diffeomorphism that takes the Poincar\'e patch a BTZ black hole. We briefly discuss the calculation of correlation functions in these backgrounds and give results at large operator dimension.
\end{abstract}

\maketitle
\tableofcontents

\section{Introduction}
The gauge-gravity correspondence \cite{Aharony:1999ti} tells us that certain quantum field theories have dual descriptions in terms of a quantum theory of gravity. Because this is a weak-strong duality, we can use semiclassical gravity to learn about strongly interacting field theories, and we can connect black hole thermodynamics with  thermodynamics of the dual field theory. In the limit where we have control over the gravitational side, we can use the correspondence to understand nontrivial behavior of systems at strong coupling. One thing in particular we can use this to study is the behavior of the system in nontrivial and possibly time-dependent excited states. By studying these states we can gain insight about the  dynamics of the strongly interacting theory.

One nontrivial way of probing the state is by studying the entanglement entropy of a spacial region $A$ with it's complement $B=\bar{A}$. This is defined by the von Neumann entropy of the reduced density matrix  $S_A=-\tr (\rho_A \ln \rho_A)$, where $\rho_A=\tr_B\rho$. The conjectured holographic prescription for calculating entanglement entropy for a static configuration \cite{Ryu:2006bv} involves finding minimal surfaces $\Sigma$ in the bulk geometry whose boundary coincides with the boundary of $A$, $\partial A =\partial \Sigma$. The entanglement entropy is then\footnote{We are working in Einstein frame with the action normalized as $S=\frac{1}{16\pi G_N}\int\sqrt{-g}R+\ldots$ and ignoring higher-derivative corrections.}
\be
S_A=\frac{1}{4G_N} \min_{\Sigma}\left[ \mathrm{Vol}(\Sigma)\right]\label{entanglement_formula}
\ee
This is only a conjectured formula, though it has passed many nontrivial checks (see e.g. \cite{Nishioka:2009un,Headrick:2010zt} for a review). When the system in question is no longer static, this must be generalized to the more nontrivial covariant holographic entanglement formula \cite{Hubeny:2007xt}. However, it was argued  that in three-dimensional gravity without extra matter fields, the fact that any solution is in fact $AdS_3$ (up to quotients) greatly simplifies calculating the entanglement entropy. We will end up using this fact to greatly simplify our calculations. 

The goal of this paper is to calculate the entanglement entropy of a conformal field theory as a pulse of energy passes through it by using holographic techniques. 
\begin{figure}[ht!]
\begin{center}
\includegraphics[scale=.45]{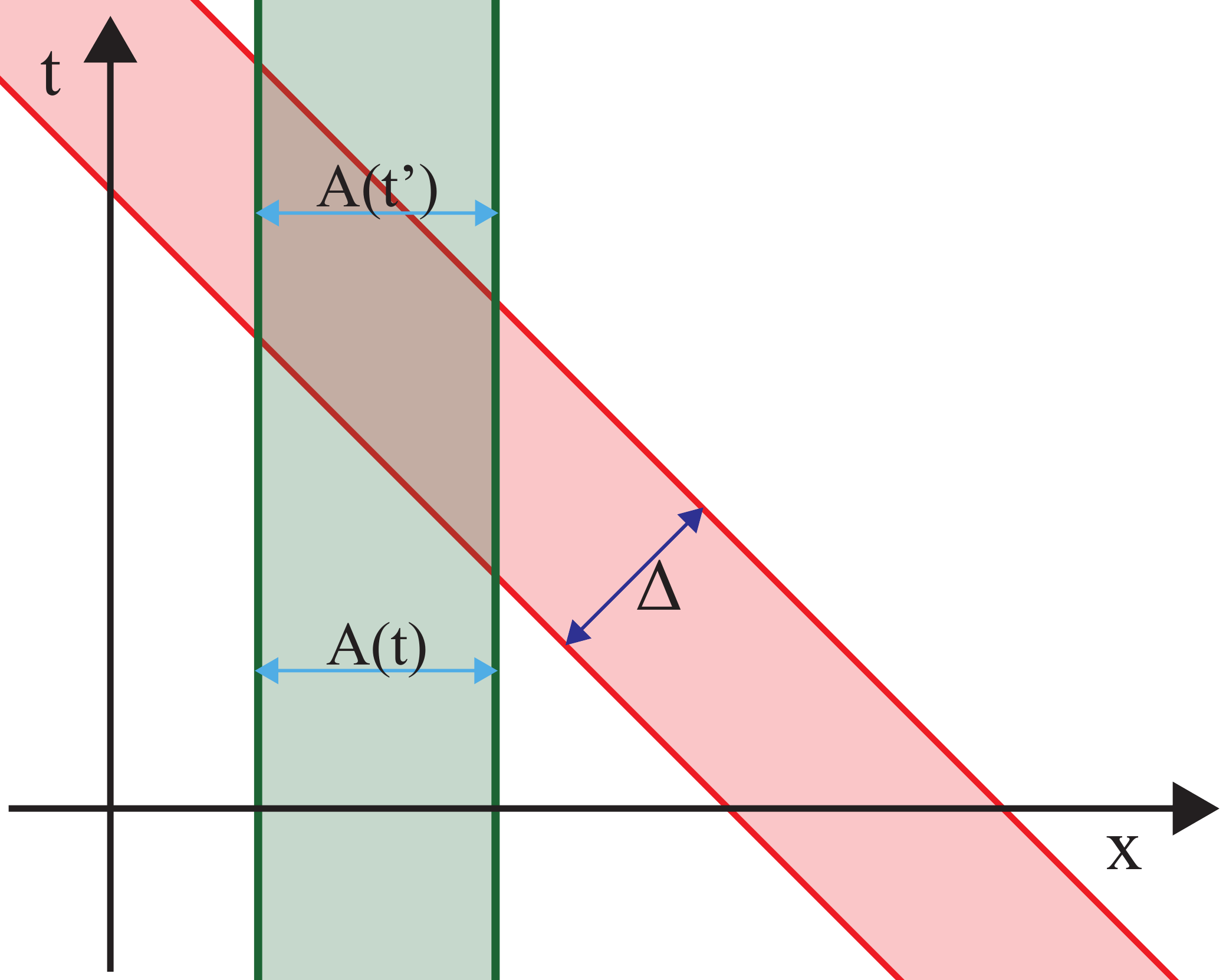}
\end{center}
\vspace{-.5cm}
\caption{\label{diagram_idea}A representation of the system we wish to construct. The red shaded diagonal area is the region of nonzero left-moving energy density with a (lightcone coordinate) width $\Delta$, and the green vertical region is the time-swept path of the spacial regions we wish to calculate entanglement entropy for.}
\end{figure}
 We demonstrate the construction in figure \ref{diagram_idea}. This problem is significantly more tractable in $AdS_3/CFT_2$ by taking advantage of the Weyl anomaly in the stress tensor transformation. It is also comparable to the time dependent entanglement entropy in a CFT subjected to a local quench \cite{Calabrese:2007fk,Asplund:2011cq}. In section \ref{section_diffs} we will review the enhanced conformal symmetry in asymptotically $AdS_3$ gravity. We will construct a finite diffeomorphism that takes us from the Poincar\'e patch to a general solution with arbitrary $T_{\pm \pm}$. In section \ref{section_wavepacket} we use this diffeomorphism to construct a solution corresponding to a left-moving packet of energy with compact support. Using the explicit diffeomorphism we are able to calculate the time dependent  entanglement entropy of a region as the wavepacket passes through it. In section \ref{section_shockwave} we take the narrow width limit of the wavepacket, reducing it to a three-dimensional Aichelburg-Sexl shockwave, and argue that this result is related to the time evolution of entanglement entropy after a local quench. In section \ref{section_CFT} we find that the transformation law derived holographically can also be found by field theory considerations, giving a nontrivial verification of the covariant holographic entanglement entropy formulation \cite{Hubeny:2007xt}. In section \ref{section_conclusions} we discuss implications and various generalizations of our results.

\section{1+1 Conformal Symmetry and $AdS_3$}\label{section_diffs}

Consider general relativity in 2+1 dimensions with a negative cosmological constant in a spacetime that's asymptotically $AdS_3$. Thanks to the fact that there are zero degrees of freedom in 2+1 GR, the most general solution to Einstein's equations is (up to gauge transformations) \cite{Banados:1998gg}\footnote{The paper of Ba\~nados works primarily in Euclidean signature but it is simple to extend the results to Lorentzian signature.}
\be
ds^2=\ell^2\left(L_+dx_+^2+L_-dx_-^2-\left(\frac{2}{z^2}+\frac{z^2}{2}L_+L_-\right)dx_+dx_- +\frac{dz^2}{z^2}\right)\label{Lmetric}
\ee
where $L_\pm = L_\pm(y_\pm)$ and the holographic stress tensor \cite{Balasubramanian:1999re} is
\be
T_{\pm\pm}=\frac{\ell}{8\pi G_N}L_\pm,~T_{\pm \mp}=0.\label{Lstress}
\ee
Of course this manifold is still (locally) exact $AdS_3$, we have just chosen a nontrivial slicing. We can in fact shift $L_\pm$ with a linearized diffeomorphism,
\be
x_\pm \rightarrow x_\pm+\delta \left( f_\pm +\frac{z^2 f_\mp''}{4-z^4 L_+L_-}+\frac{z^4L_\mp f_\pm''}{8-2z^4L_+L_-} \right),~z\rightarrow z\left(1+\delta \frac{f_+'+f_-'}{2}\right),\label{general_inf_diff}
\ee
which shifts  $L_\pm$ as
\be
\Delta L_\pm=f_\pm L_\pm'+2L_\pm f_\pm'-\frac{f_\pm'''}{2}.
\ee
Integrating this diffeomorphism to nonlinear order in a closed form is not trivial, though it has been done for the simple case of a constant $L_\pm$, which is just the change of slicing to go from $AdS_3$ to the BTZ black hole \cite{Carlip:1994gc}. We find that it is also possible to construct the nonlinear diffeomorphism when the starting coordinate slicing is that of Poincar\'e $AdS_3$. Starting with the metric
\be ds^2 =\ell^2 (-2 dy_+ dy_- + du^2)/u^2,\ee
 acting with the coordinate transformation $(y_+,y_-,y) \rightarrow (x_+,x_-,z)$ given from
 \be
 y_\pm = f_\pm(x_\pm)+\frac{2z^2f_\pm'^2f_\mp''}{8f_\pm'f_\mp'-z^2f_\pm''f_\mp''},~u=z\frac{\left(4f_+'f_-' \right)^{3/2}}{8f_+'f_-'-z^2f_+''f_-''}\label{FG_diff}, \ee
 which linearizes to (\ref{general_inf_diff}) with the original $L_\pm(y_\pm)=0$, and asymptotes to
  \be
 y_\pm = f_\pm (x_\pm)+\Op(z^2),~u=z \sqrt{f_+' f_-'}+\Op(z^3)\label{FG_asymptotic_diff},
 \ee
the metric is precisely (\ref{Lmetric}) with $L_\pm(x_\pm)$ proportional to the Schwarzian derivative of $f_\pm$,
 \be L_\pm=\frac{1}{2}\{f_\pm,x_\pm\}=\frac{3f_\pm''^2-2f_\pm'f_\pm'''}{4f_\pm'^2}. \label{L_diff_eq}\ee
We see from (\ref{Lstress}) the stress tensor transformation satisfies the general transformation law  with $T_{ab}(y_\pm)=0$. When the theory is on a cylinder, we have to be very careful about how the map changes the boundary conditions, so we will consider the theory on $\M^{1,1}$.  Thanks to our closed form diffeomorphism we know solving any problem such as wave scattering or geodesic motion on (\ref{Lmetric}) reduces to solving that same problem on Poincar\'e AdS and then following the coordinate map, often paying careful attention to what we do to the boundary. We can use (\ref{FG_diff}) to map solutions to, for instance, the Klein-gordon equation in $y$ coordinates to nontrivial $x$ coordinates. However, as we are working in Lorentzian signature, there will be nontrivial subtleties in terms of boundary conditions on the horizon if we are interested in calculating different-time correlators. Thanks to the closed-form finite diffeomorphism these can be sorted out by inspection once we solve (\ref{L_diff_eq}). 

It needs to be emphasized that there can be other states in the CFT with the same stress tensor expectation values, which may differ by the behavior of other operators. In the bulk this would correspond to turning on other supergravity fields. These other solutions will not be related to planar AdS by a simple diffeomorphism, and so we can not as easily study entanglement in these backgrounds. However, when the scale of the stress tensor expectation value is much greater than the one-point functions for other operators, this background will be a good approximation.\footnote{We thank Matthew Headrick for emphasizing this to us.}
 
\section{Left-moving ``wavepackets''}\label{section_wavepacket}
\begin{figure}[ht!]
\begin{center}
\includegraphics[scale=.85]{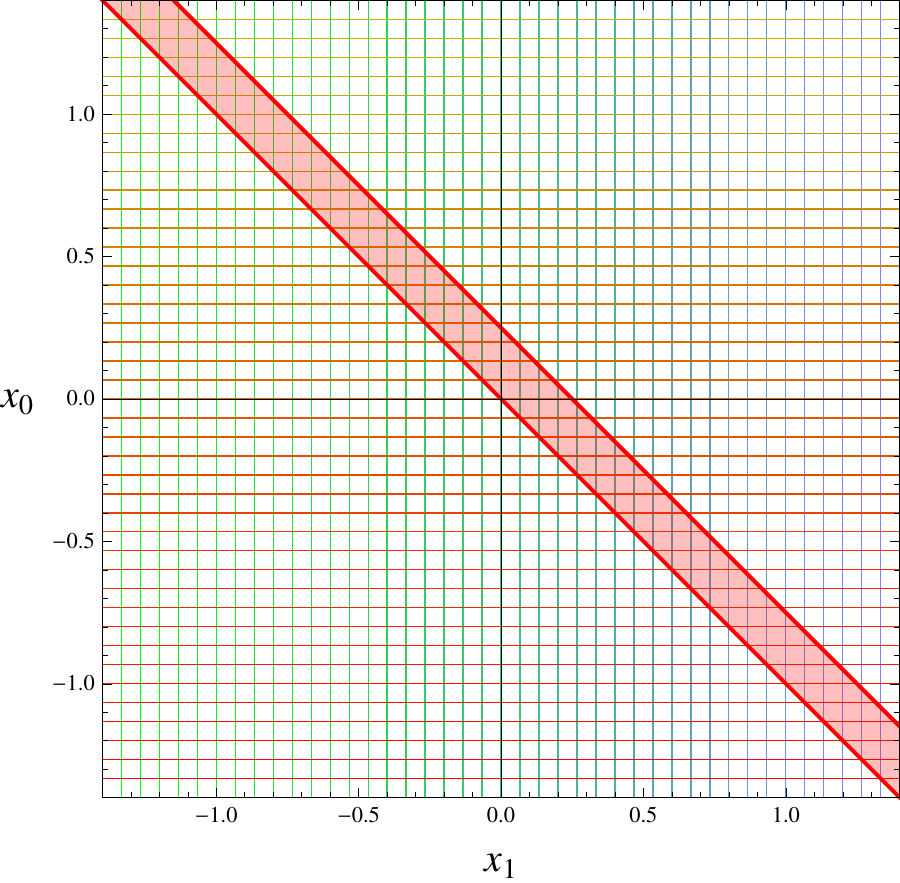}
\includegraphics[scale=.85]{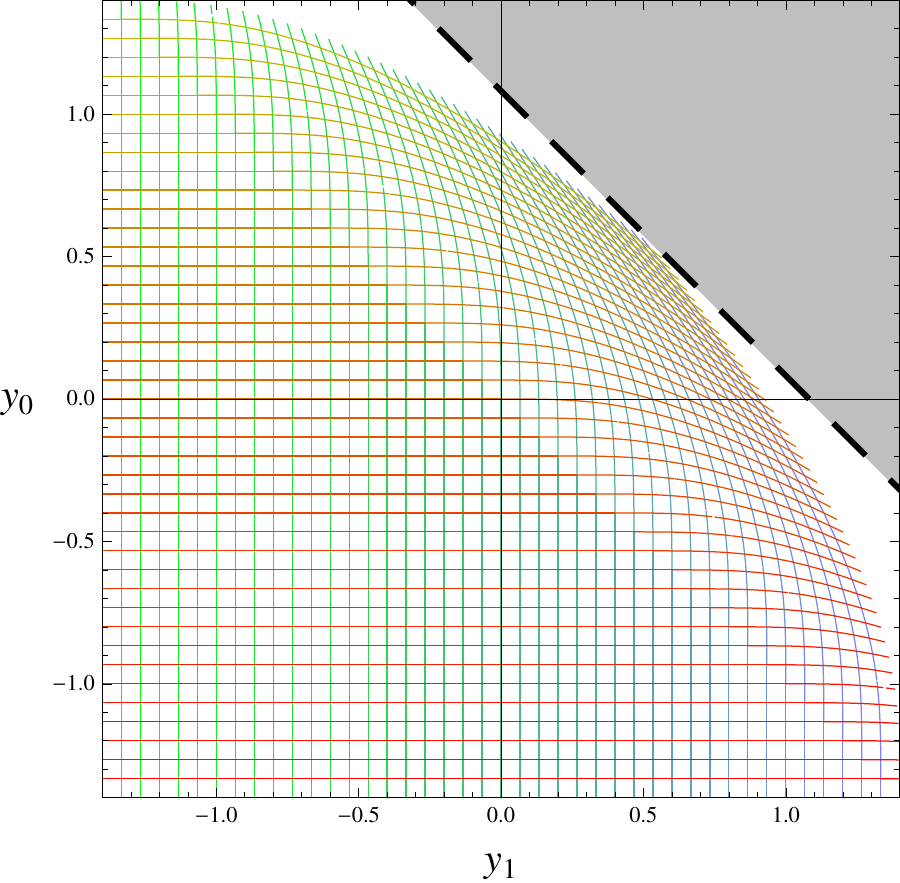}\end{center}
\vspace{-.5cm}
\caption{\label{inverse_map}The map $f_+$ for $\Delta =1/4,~ \tau=2.$ Note that the map is the identity for $x_+<0$. On the left we also show the region with nonzero $T_{++}$ and on the right we shade the region of $x$ not covered by the $y$ coordinates.}
\end{figure}

\begin{figure}[ht!]
\begin{center}
\includegraphics[scale=.75]{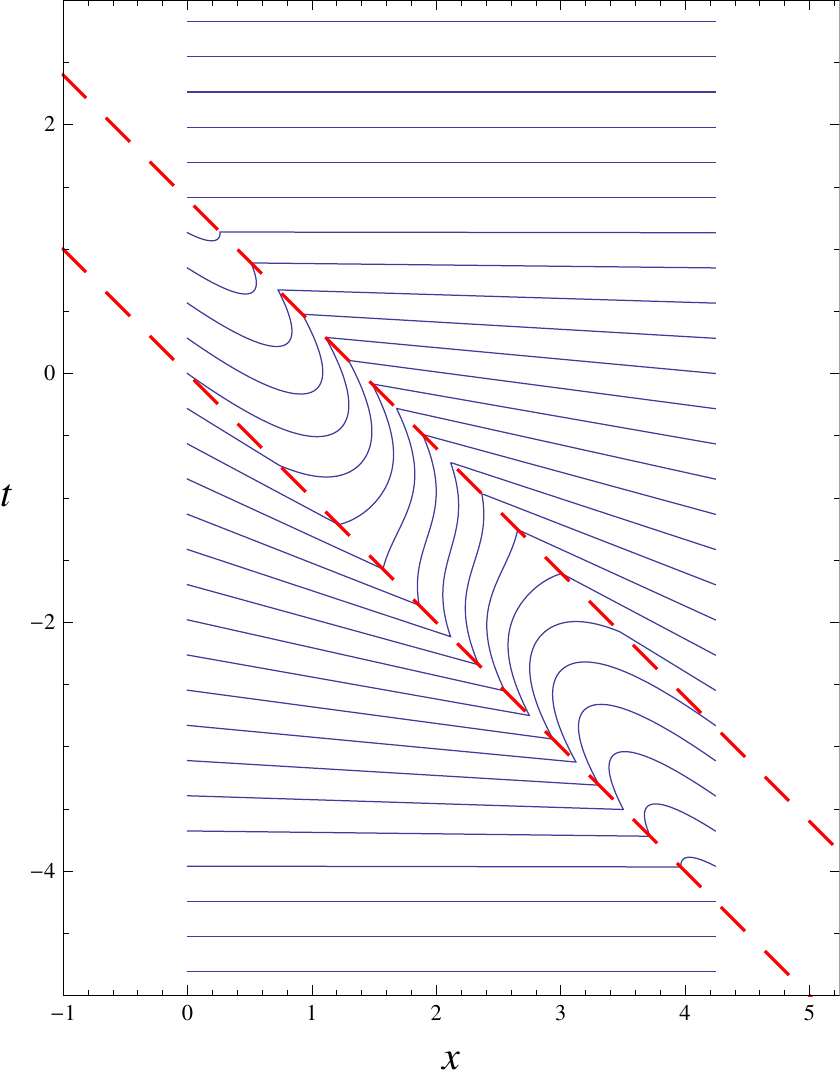}
\includegraphics[scale=.75]{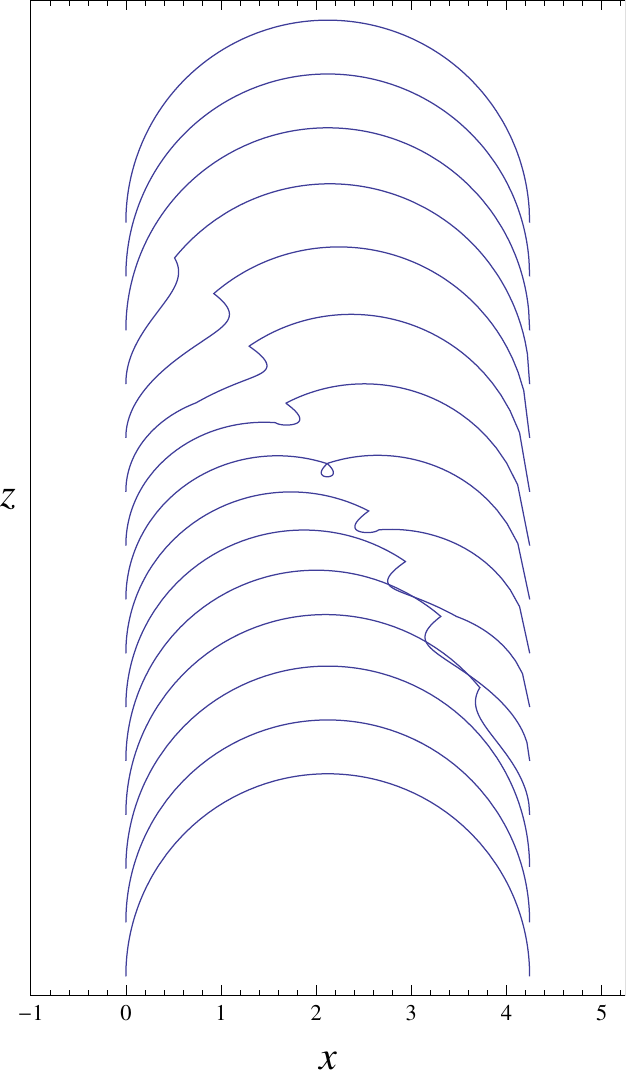} \end{center}
\vspace{-.5cm}
\caption{\label{geod} Geodesics connecting two points at equal $x$ time, with $\tau=2,~\Delta=1,~d=3$. On the left we have the projection to the $(x,t)$ plane, with dashed red lines denoting where the region of nonzero $L_+$ is.  On the right we have the projection onto the $(x,z)$ plane, with $z$ shifted by time for clarity. Note that while the curves look timelike, this is due to the frame-dragging from $L_+$ and they are indeed spacelike. Note that when $\Delta > d$ the curves are very similar, and when they are entirely within the pulse they are simply extermal BTZ geodesics.}
\end{figure}

Let's start by constructing a simple left-moving wavepacket with a constant nonzero $T_{++}$ for $0<x_+<\Delta,$ and $L_-=0$ everywhere. We can think of this as an $AdS_3$ version of an AdS Aichelburg-Sexl shockwave \cite{Hotta:1992qy}, where the stress tensor is spread over a finite region of $x_+$ and not just delta-function supported. The simplest model is one where
\be
L_+=\left\{
\begin{array}{ccc}
0 & | & x_+<0 \\
\tau^2 & | & 0\le x_+ \le \Delta \\
0 & | & \Delta < x_+
\end{array}\right.,~ L_-=0.\label{L_step}
\ee
This is only piecewise smooth, but we know that we could easily replace $L_+$ with a function that is smooth and still has compact support. We will find that this discontinuity in $L_+$ means that the entanglement entropy has a discontinuous second derivative in $x_+$. We now simply need to solve (\ref{L_diff_eq}) with the source given by (\ref{L_step}). First, let's point out the diffs that give $L_+=0$ are $f_+ = (a+b c_+)/(c+d c_+)$ and similarly for the right-moving sector\footnote{This is just the global $SL(2,\mathbb{R})\times SL(2,\mathbb{R})$ symmetry.}. Because we are not exciting the right-moving sector, we can happily pick $f_-=x_-$.  For $x_+<0 $ we will simply use $f_+=x_+$ and a general $SL(2,\mathbb{R})$ element for  $x_+>\Delta$. What about a positive constant $L_+=\tau^2$? A general solution is
\be
f_+=a+b\tanh\left[\tau x_++c \right].
\ee
We need $T_{++}$ to be piecewise continuous, so matching $f_+,~f_+',~f_+''$ at $x_+=0$ and $x_+=\Delta$ we find
\be
y_+=f_+(x_+)=\left\{\begin{array}{ccc}
x_+ & | & x_+<0 \\
\tanh[\tau x_+]/\tau & | & 0\le x_+ \le \Delta \\
\frac{1}{\tau}\frac{1+(x_+-\Delta)\tau\coth[\tau\Delta]}{(x_+-\Delta)\tau+\coth[\tau\Delta]} & | & x_+>\Delta
\end{array}\right.\label{diff_function}
\ee
We demonstrate this map in figure \ref{inverse_map}.
Note that the map does not cover all of $y_+$, only $y_+ < \coth[\tau \Delta]/\tau$. This is similar to what happens if we act with a diffeomorphism that takes us to an extreme left-moving BTZ black hole, where $f_+=\exp(\sqrt{L_+}x_+)$. We know that geodesics in $y$ coordinates connecting two points on opposite sides of this boundary must map to a geodesic which begins on the $x$ boundary and crosses a horizon in the interior \cite{Maldacena:2004rf}\footnote{It may seem as though acting with a global $SL(2,\bbR)$ rotation might alleviate this, but since an element of $SL(2,R)$ maps $\mathbb{R} \bigcup \{\infty\} \mapsto\mathbb{R} \bigcup \{\infty\} $, we know even with this rotation the range of this function will never cover all $\bbR$.}. However, since $f_+$ is invertible a geodesic connecting two spacelike-separated boundary points in $x$ coordinates will map to a spacelike geodesic connecting two boundary points with both $y_+<\coth[\tau\Delta]/\tau$. We can demonstrate this by using $f_+$, its inverse, and boosts to map geodesics connecting points at equal $y_\pm$ time, $u(y)=\sqrt{(y-y_1)(y_2-y)}$, to geodesics connecting two equal $x_\pm$ time points in $x$ coordinates. The explicit functional form of the geodesics is messy and not terribly enlightening, so we provide plots to help visualize in figure \ref{geod}.

Now we wish to measure the entanglement entropy for a region $A$ between two points in $x_\pm,z$ coordinates. This is related to the (regulated) AdS geodesic distance between the two points by simply mapping the problem to $y$ coordinates. This is quite simple, as in $y_\pm$ coordinates the answer (for spacelike seperated points) is
 \be
S_{ent.} (y_1,y_2)=\frac{c}{12}\log\left[\frac{4 (y_{2+}-y_{1+})^2 ( y_{2-}-y_{1-})^2}{\epsilon_2^2~\epsilon_1^2} \right],~c=3\ell/2 G_N.\label{single_entanglement}
 \ee
 where $\epsilon_i$ is the location of the radial cutoff $u_i$, assumed to be much smaller than any other bulk length scale. It is important to keep track of these, as our answer in terms of $x$ coordinates is meant to be asked with the same regulator on each point in $z$, not $u$. Note that if we calculated an object which is UV finite such as a mutual information, these terms will cancel but it is possible that subleading terms in (\ref{FG_diff}) may be relevant, which would indicate some regularization-scheme dependence of the result. 
 
\begin{figure}[ht!]
\begin{center}
\includegraphics[scale=.5]{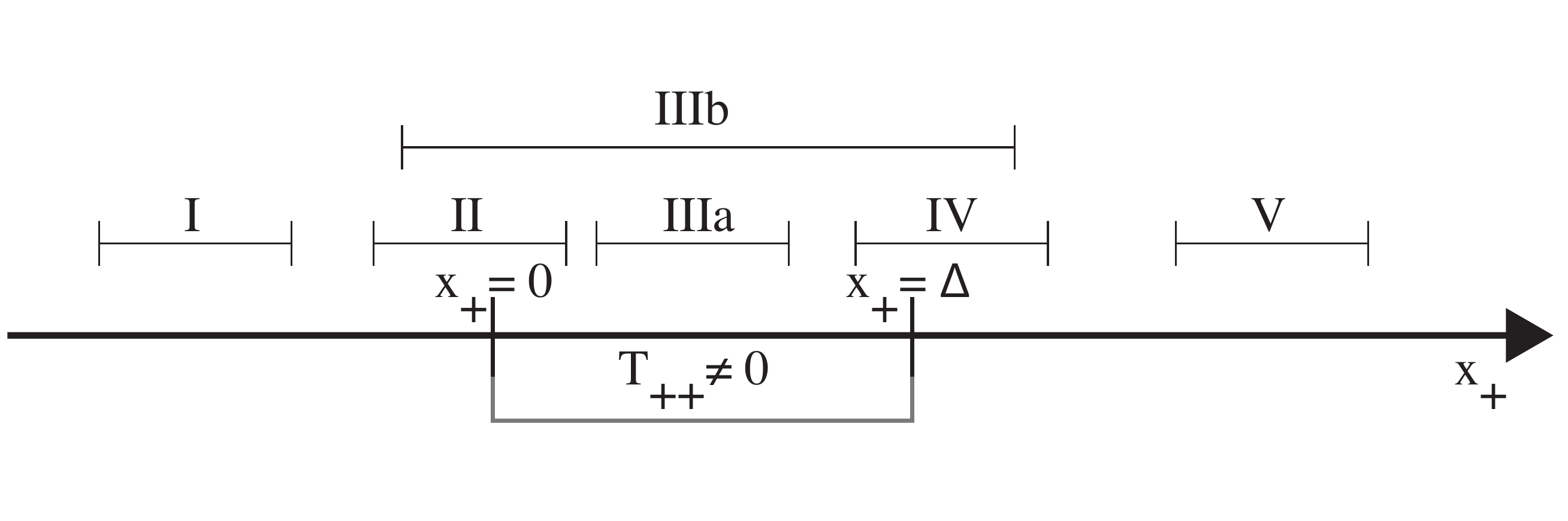}\end{center}
\vspace{-.5cm}
\caption{\label{regions}The six regions for the piecewise smooth function (\ref{D_regions}), labelled I - V. The region of interest is always between $x_+-d$ and $x_+$. Note that IIIa only occurs for $d<\Delta$ and IIIb for $\Delta<d$.}
\end{figure}

\begin{figure}[ht!]
\begin{center}
\includegraphics[scale=.85]{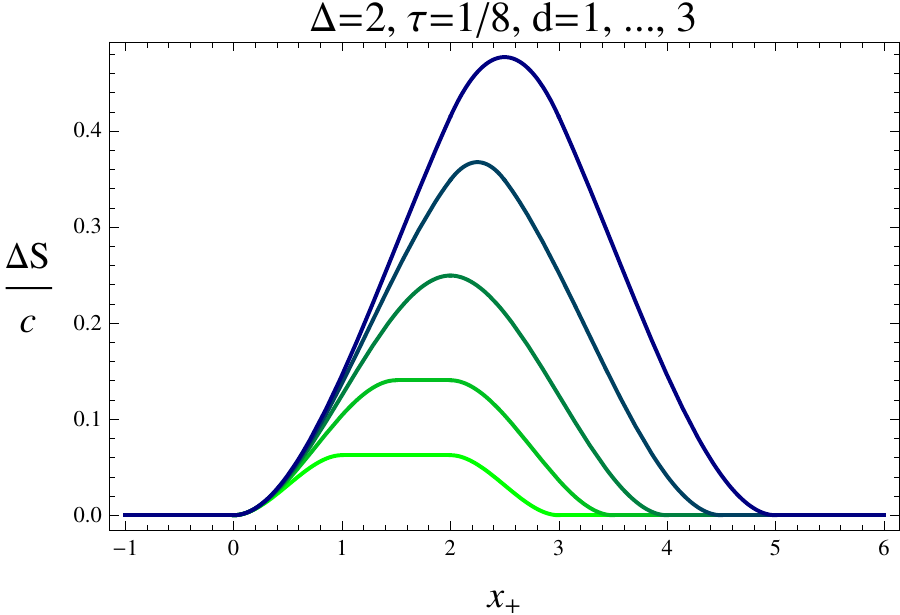}
\includegraphics[scale=.85]{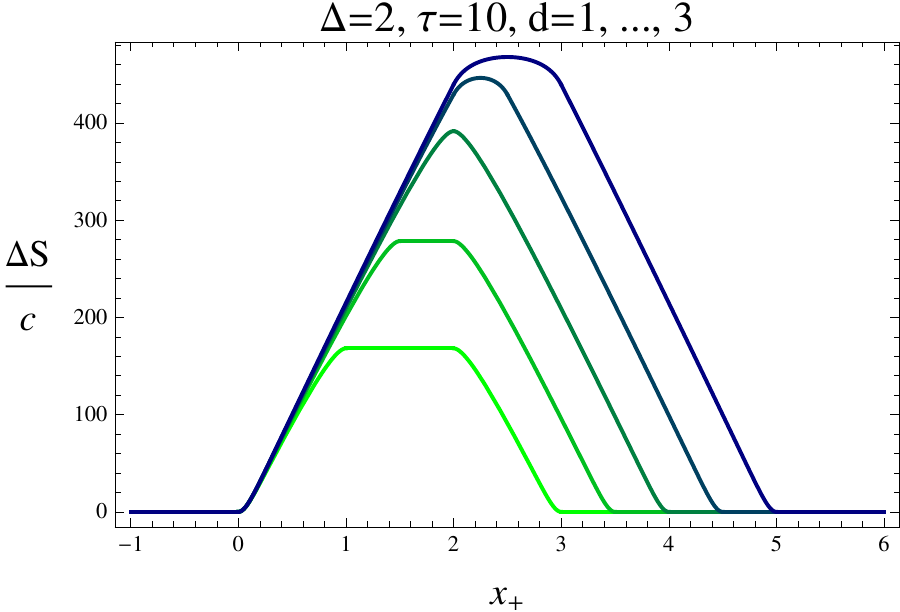}\end{center}
\vspace{-.5cm}
\caption{\label{entanglement} Plots of $\Delta S= S-\frac{c}{6}\log\left[ 2d^2/\delta^2\right]$ at fixed $\tau$ and $\Delta$ as we increase $d$.  The lightest curve has $d=1<\Delta$ and the darkest curve has $d=3>\Delta$.}
\end{figure}

 Let's first work out how (\ref{single_entanglement}) transforms under a general left-moving transformation $f_+$ with $f_-(x_-)=x_-$. The result is
 \be
  S_{ent.}(x_1,x_2)=\frac{c}{12}\log\left[\frac{4(f_+(x_{2+})-f_+(x_{1+}))^2(x_{2-}-x_{1-})^2}{f'_+( x_{2+})f'_+(x_{1+})\delta_1^2\delta_2^2}\right]\label{general_diff_entangle}
 \ee
 where $\delta_i$ is the radial cutoff in $z_i$ and we have used the fact that we require 
 \be
\delta\ll~(\mathrm{scales~in}~f_+).
 \ee
 This seems like it should generalize when $f_-$ is nontrivial as well to
\be
  S_{ent.}(x_1,x_2)=\frac{c}{12}\log\left[\frac{4[f_+(x_{2+})-f_+(x_{1+})]^2[f_-(x_{2-})-f_-(x_{1-})]^2}{f'_+( x_{2+})f'_+(x_{1+})f'_-( x_{2-})f'_-(x_{1-})\delta_1^2\delta_2^2}\right]\label{left_right_diff}
 \ee 
 which would be true as long as we can verify that the connected geodesic between the two points is always shorter than a piecewise smooth geodesic which falls down into the horizon and runs along it. However we will find from CFT considerations that this is precisely the behavior we find. Recall that when $g \in SL(2,\mathbb{R})$, 
 \be
\frac{ (g(x)-g(y))^2}{g'( x) g'(y)} =(x-y)^2
 \ee
 and so despite  swirling the coordinates around, the regulated geodesic distance between two boundary points is invariant. This  means that our entanglement entropy will go back to the vacuum value when the pulse has completely passed through the region, despite the coordinate transformation $f_+$ no longer being simply the identity. Indeed, plugging (\ref{diff_function}) into  (\ref{general_diff_entangle}) and evaluating at equal times,
 \be
t_1=t_2,~x_1=x_2+\sqrt 2 d ,~\mathrm{i.e.}~x_{1\pm}=x_{2\pm}\pm d
 \ee
we find
 \be
 S_A = \frac{c}{6} \log \left[2  D^2/\delta^2 \right],~\Delta S_A = S_A - S_{vac.} = \frac{c}{6}\log[D^2/d^2],
 \ee

\be
D^2=\left\{\begin{array}{ccc}
d^2 & | & x_+<0 ~ (I) \\
d\cosh[\tau x_+](d+\tanh[\tau x_+]/\tau-x_+) & | & 0\le x_+\le \Delta,~x_+<d ~(II)\\
d\sinh[\tau d]/\tau & | & 0\le x_+\le \Delta,  0 \le x_+-d\le \Delta~(IIIa)\\
d\cosh[\tau\Delta]& | & \Delta<x_+,~x_+-d<0~(IIIb)\\
\times \left(d-\Delta+\frac{(1+(d-x_+)(x_+-\Delta)\tau^2)\tanh[\tau\Delta]}{\tau}\right)  &  &\\
d\cosh[\tau(d+\Delta-x_+)]& | & \Delta<x_+,\Delta\le x_+-d\le 0~(IV) \\
\times\left(x_+-\Delta+\frac{\tanh[\tau(d+\Delta-x_+)]}{\tau} \right) &  &\\
d^2 & | & \Delta+d<x_+ ~ (V)
\end{array}\right.\label{D_regions}
\ee
 To clarify this we include a figure showing what regions we are considering in figure \ref{regions}.
Note that $D^4$ is symmetric under a reflection about $x_+=\frac{d+\Delta}{2}$, where it has a maximum. This gives us the peak entanglement entropy
\be
\max(\Delta S)=\left\{\begin{array}{ccc}
\frac{c}{6}\log\left[ \frac{\sinh[\tau d]}{\tau d} \right] & | & d < \Delta\\
\frac{c}{6}\log\left[ \frac{4(d-\Delta)\tau\cosh[\tau\Delta]+(4+(d-\Delta)^2\tau^2)\sinh[\tau\Delta]}{4 \tau d} \right] & | & \Delta \le d
\end{array}\right.\label{Smax}
\ee
We also provide plots of $\Delta S$ for various cases in figure \ref{entanglement}. We again emphasize that this is the entanglement entropy for the state created by acting with the diffeomorphism (\ref{diff_function}) on the vacuum, and it is possible that there are other states with the same stress tensor profiles but with for instance nonvanishing one-point functions for other operators. However for states where the stress tensor scale is much greater than any other scales, the metric we use is a reasonable approximation, and the resulting entanglement entropy should be similar. In bulk language, we expect that as long as other fields are not turned on too strong and so do not backreact too much on the metric, this result should be a very good approximation.

Note that in the region IIIa the entanglement entropy agrees with the entanglement entropy for an extremal BTZ black hole. One can extend our work to easily show that if we have left- and right-moving packets and we measure the entanglement entropy using (\ref{left_right_diff}) in the overlap region we find the entanglement agrees with the nonextreme BTZ black hole \cite{Hubeny:2007xt},
\be
S_{overlap}=\frac{c}{6}\log\left[ \frac{2\sinh[\tau_+ d]\sinh[\tau_- d]}{\tau_+ \tau_-\delta^2} \right],~T_{\pm\pm}=\frac{\tau_\pm^2\ell}{8\pi G_N}.
\ee
However, the overlap region does not extend to late times, which means that the system does not actaully thermalize. This is because we are only considering vacuum gravity in 2+1, and so the left and right moving sectors do not interact.

Thanks to the fact that we know the geometry is exact $AdS$, we can use the arguments of \cite{Hubeny:2007xt} to know that we are in fact calculating the covariantly defined holographic entanglement entropy. It is worth pointing out that the only other possible geodesic connecting the two points is a piecewise smooth geodesic which falls down to the horizon from the two points, and then runs along it. We know from the form of our metric (\ref{Lmetric}) that the horizon sits at $z^4=4(L_+ L_-)^{1/4},$ and, thanks to the fact that we only have left-moving stress tensor turned on the horizon sits at $z=0$ and a geodesic falling down will always have an infrared divergent length, and therefore the shortest geodesic is the connected one whose length we have used.

\section{Shockwave limit}\label{section_shockwave}
\begin{figure}[ht!]
\begin{center}
\includegraphics[scale=.85]{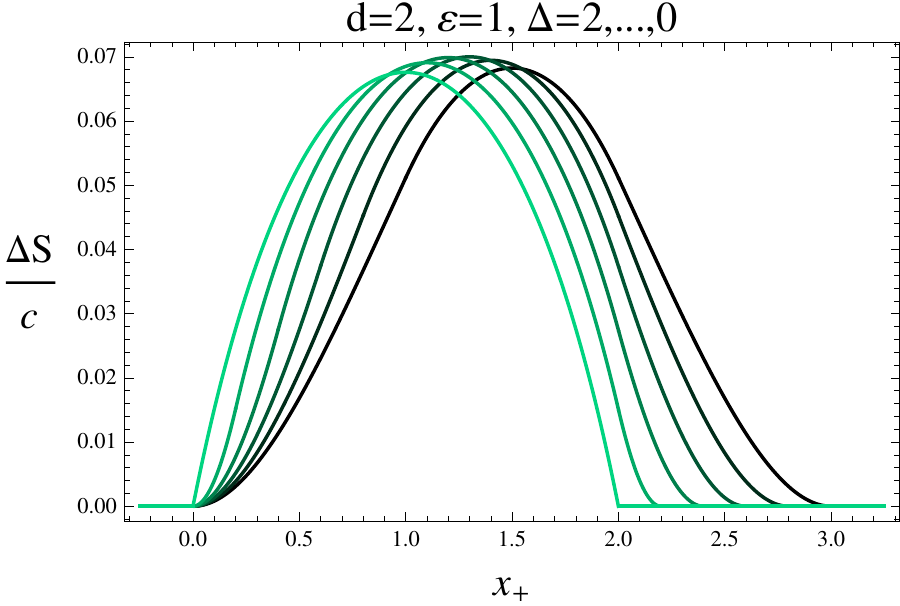}
\includegraphics[scale=.85]{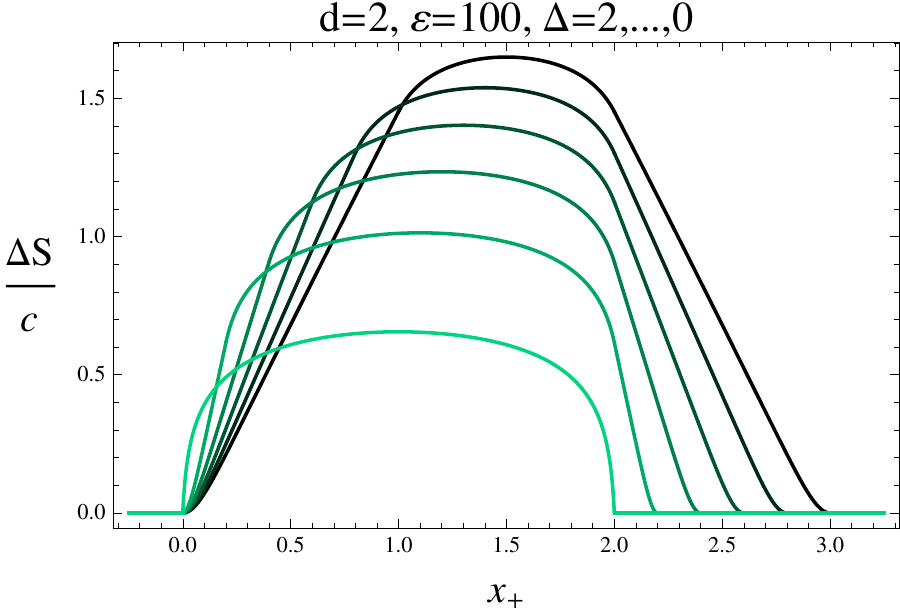}\end{center}
\vspace{-.5cm} 
\caption{\label{pulse_limit} Plots of $\Delta S$ at fixed $d$ and $\varepsilon$ as we send $\Delta \rightarrow 0$. The darkest curve is $\Delta=2$ and it lightens as we approach $\Delta=0$.}
\end{figure}

Now we will discuss what happens when we try to make our pulse of energy look more like the traditional AdS Aichelburg-Sexl shockwave \cite{Hotta:1992qy} (for a review see eg \cite{Gubser:2010nc}). Recall that the shockwave metric in $AdS_{d+1}$ is
\be
ds^2=\frac{\ell^2}{z^2}\left[ -2 dx_+ d_- + d\vec{x}_T^2 + dz^2 \right]+ \Phi(x_T,z) \delta(x_+) dx_+^2\label{genlshock}
\ee
where $\Phi$ is a nontrivial function. We see that our piecewise geometry is a simple example of a shockwave, but since we do not have transverse directions in $AdS_3$ we do not \emph{have} to support all of the stress tensor on a delta function. Noting that the total energy on a constant $t$ slice is
\be
E=\int T^{00} ~ dx =\int  \frac{\ell}{16\pi G_N}L_+\left[\frac{t+x}{2^{1/2}}\right] dx= \frac{\ell}{16\pi G_N} \sqrt{2}\Delta\tau^2.
\ee
Keeping $\varepsilon = \Delta \tau^2$ finite while sending $\Delta \rightarrow 0$ simplifies things, as now the pulse is localized solely at $x_+=0$. Comparing (\ref{genlshock}) to our metric we find $\Phi =\ell^2\sqrt{2}\varepsilon$, which agrees with the $d=2$ shockwave metric. The entanglement entropy reduces to 
\be
\Delta S=\left\{\begin{array}{ccc}
0 & | & x_+<0  \\
\frac{c}{6}\log\left[1+\frac{\varepsilon x_+(d-x_+)}{d} \right]& | & 0<x_+,~x_+-d<0\\
0 & | & d<x_+ 
\end{array}\right.\label{S_pulse}
\ee
Note that in this limit the complicated behavior in the entanglement has simplified dramatically, and we only have an entanglement entropy different from the vacuum value when the pulse is within the region $A$. The maximum entanglement entropy (\ref{Smax}), now occuring at $x_+=d/2$,  has simplified to
\be
\max[\Delta S]=\frac{c}{6}\log[1+d \varepsilon/4],
\ee
which features logarithmic scaling for large $d\varepsilon$. Since $L_+$ is now a delta-function and not a piecewise-continuous, the entanglement entropy has a discontinuous first derivative. In figure \ref{pulse_limit} we provide plots of $\Delta S$ as we send $\Delta \rightarrow 0$, with fixed $\varepsilon$.

This result is very similar to CFT calculations of entanglement entropy of 1+1 systems subjected to local quenches, which are constructed by adding or removing a point-like defect seperating a spacial slice into two regions \cite{Calabrese:2007fk, Asplund:2011cq}. Changing the topology of the spacetime will cause a local UV divergence in the form of an infinite-energy pulse leaving the defect at the speed of light. If we join two manifolds separated by a co-dimension one boundary with transverse translational symmetry, it is reasonable to expect that this will source a bulk gravitational shockwave, which we can model as an Aichelburg-Sexl shockwave. Indeed, it the limit $\varepsilon \gg 1$, we find (\ref{S_pulse}) looks very similar to the CFT results, where we identify $\varepsilon$ with the UV regulator of the local quench, and we find that there is no residual entanglement entropy after the pulse has passed, indicating we have lost information of the boundary entanglement entropy. This one missing component can be included by assuming there is a contribution to the initial entanglement entropy coming from the co-dimension one boundary defect, treating it perhaps along the lines of \cite{Takayanagi:2011zk}.

The entanglement entropy for the $AdS_3$ shockwave can be generalized to higher dimensions. Further we can use the fact that it can be constructed by taking a singular limit of a boosted global AdS black hole. We can then use the covariant holographic entanglement entropy construction and apply it to nontrivial time-depentent boundary regions and find a higher-dimensional analog of (\ref{S_pulse}). While we leave this calculation to future work, we conjecture that in general the form will be similar, with $\max[\Delta S] \propto (d \varepsilon)^{D-2}$ as $d\varepsilon \gg 1$. where we are working with a $D$-dimensional CFT, $d$ is the length scale of the boundary region $A$, and $\varepsilon$ is the energy of the shockwave.

\section{CFT comparison}\label{section_CFT}
Calculating an entanglement or R\'enyi entropy in a generic state in a CFT is quite nontrivial. However when we consider states constructed by acting with a nontrivial conformal transformation on the vacuum, we can use the fact that the replica trick involves twist operators which transform as primaries under conformal transformations \cite{Calabrese:2004eu}. This argument has been used previously to calculate the entanglement entropy for a thermal state or the theory on a circle. If we can analytically continue the argument of \cite{Calabrese:2004eu} to Lorentzian signature, we expect that $\tr \rho_A^n$ transforms under scale and conformal transformations as the $n$th power of a primary operator $\Op_n$ with $\Delta_+=\Delta_-=\frac{c}{24}(1-1/n^2)\equiv \Delta_n$, where $\Delta_\pm$ are now left and right central charges. This is a nontrivial assumption, because the formula for $\tr \rho_A^n$ is derived from a Euclidean path integral, and we are considering states which can not be represented as Euclidean path integrals. In vacuum, for a single interval,
\be
\tr \rho_A^n = c_n \left|\frac{2(y_{2+}-y_{1+})(y_{1-}-y_{2-})}{a}\right|^{-\frac{c}{12}\left(n-1/n \right)} = \langle \Op_n (y_1)\Op_n(y_2)\rangle^n.\ee
where $c_n$ are not generically determined, except for $c_1=1$, and $a$ is a nonuniversal UV lattice scale. Recall that under the map $x_\pm \mapsto y_\pm=f_\pm(x_\pm)$,
\be
\langle\Op_n(x_1)\Op_n(x_2)\rangle^n =\left| f_+'(x_{1+}) f_-'(x_{1-}) f_+'(x_{2+}) f_-'(x_{2-}) \right|^{ n  \Delta_N}\nonumber\ee
\be
\times \left| \frac{2\left[ f_+(x_{2+})-f_+(x_{1+})\right]\left[ f_-(x_{1-})-f_-(x_{2-})\right]}{a^2} \right|^{-2n \Delta_n}
\ee
We can now evaluate $S_A =  - \partial_n\left[ \tr \rho_A^n \right]|_{n=1}$ and we find exactly (\ref{left_right_diff}),
\be
S_A=\frac{c}{6}\log\left|\frac{2\Delta f_+(x_{1,2}) \Delta f_-(x_{1,2})}{a^2\sqrt{f_{+2}'f_{+1}'f_{-1}'f_{-2}'}}\right|.\label{CFT_trans}
\ee
We wish to emphasize that since this is so far just an analytic continuation of the behavior of euclidean path integrals, (\ref{CFT_trans}) is only a conjecture for generic coordinate transformations. While one would like to simply say this is a consequence of conformal invariance of the theory, the statement is not quite so straight-forward. The important $f_\pm'$ contribution to the entanglement came from very different places on either side of the duality. On the field theory side, it is a  consequence of the fact that the twist operators transformed as primaries. However, on the gravitational side, it came about by carefully treating the transformation of the UV regulation of the geodesic length. The simplicity with which this worked out suggests that transformation laws for more complicated objects, such as mutual R\'enyi information, or mutual entanglement information, for multiple disconnected regions may have similar transformation properties. It would be very interesting to better understand  holographic R\'enyi entropies to verify that they satisfy the transformation properties given above.

\section{Conclusions}\label{section_conclusions}
In this work, we have calculated the time evolution of the entanglement entropy of a  1+1 CFT subjected to a localized packet of energy density. As a tool in doing this calculation we have integrated the infinitesimal coordinate transformations that shift the stress tensor of the CFT, providing a closed-form diffeomorphism that takes $AdS_3$ in the Poincar\'e patch to the most general solution. We have also verified that our result agrees with a CFT calculation of entanglement entropy of a single interval for a state created by acting on the vacuum with a coordinate transformation. We find that in the narrow shockwave limit, the form of the entanglement entropy simplifies greatly, and reproduces some but not all of the structure of CFT calculations of time evolution of entanglement entropy using CFT techniques \cite{Calabrese:2007fk, Asplund:2011cq}.

There are many interesting future directions one may wish to look towards related to this work. First of all, the ability to construct nontrivial nonperturbative time-dependent backgrounds in asymptotically $AdS_3$ is a powerful tool. It would be interesting to study general retarded correlators  on (\ref{Lmetric}) by mapping general solutions to the Poincar\'e patch slicing. We can use the WKB approximation to relate the bulk geodesic length between two space-like separated points to a two-point function for high dimension operators \cite{Balasubramanian:1999zv,Louko:2000tp}, so identifying $G(x_1|x_2) \approx e^{-m\times \mathrm{Length}(\gamma_{x_1,x_2})}$ with $m=\Delta/\ell+\ldots$, we can generalize (\ref{D_regions}). In the shockwave limit we find
\be
G(x_\pm | x_\pm \mp \Delta x_\pm)=\delta^{2\Delta}
\left\{\begin{array}{ccc}
\left(2\Delta x_+ \Delta x_- \right)^{-\Delta} & | & x_++\Delta x_+<0 \\
\left(2(\Delta x_+-x_+(x_++\Delta x_+)\varepsilon)\Delta x_- \right)^{-\Delta} & | & x_+<0<x_++\Delta x_+ \\
\left(2\Delta x_+ \Delta x_- \right)^{-\Delta}  & | & 0<x_+
\end{array}\right.\ee
where we have assumed that $\Delta x_\pm >0$, ensuring that the points are spacelike separated. It would be very interesting to extend this result to the two-point function for a classical field in the shockwave spacetime. Similarly it would be very interesting to use the singular diffeomorphism which gives us the higher dimensional shockwave metric (\ref{genlshock}) and study nontrivial time-dependence of correlators there.

It would also be very interesting to study various methods of quenching a holographic system which would source the shockwaves we consider, such as inserting or removing a holographic defect \cite{Karch:2001cw,DeWolfe:2001pq} or a boundary \cite{Takayanagi:2011zk}. It would also be interesting to study the inclusion of nontrivial behavior of other primary operators either in the field theory directly or as additional bulk fields, and see if with their inclusion the system could thermalize. Whether we could continue to use purely analytic techniques or would need to resort to numerical methods is unclear.
 
\vskip .5cm

\centerline{\bf Acknowledgements}
\vskip .5 cm
We would like to thank Curtis Asplund, Matthew Headrick, Matthew Kleban, and Massimo Porrati for useful discussions. We would also like to thank Matthew Headrick for comments on an early version of this draft. M.M.R. is supported by the Simons Postdoctoral Fellowship Program.
\eject
\bibliographystyle{utphys}
\bibliography{dlrrefs}

\end{document}